# Control of interlayer excitons in two-dimensional van der Waals heterostructures

Alberto Ciarrocchi[1,2,†], Dmitrii Unuchek[1,2,†], Ahmet Avsar[1,2], Kenji Watanabe[3], Takashi Taniguchi[3], Andras Kis[1,2,*]

[1]Electrical Engineering Institute, École Polytechnique Fédérale de Lausanne (EPFL), CH-1015 Lausanne, Switzerland
[2]Institute of Materials Science and Engineering, École Polytechnique Fédérale de Lausanne (EPFL), CH-1015 Lausanne, Switzerland
[3]National Institute for Materials Science, 1-1 Namiki, Tsukuba 305-0044, Japan

[†]These authors contributed equally
[*]Correspondence should be addressed to: Andras Kis, andras.kis@epfl.ch

**Long-lived interlayer excitons with distinct spin-valley physics in van der Waals heterostructures based on transition metal dichalcogenides make them promising for information processing in next-generation devices[1]. While the emission characteristics of interlayer excitons in different types of hetero stacks have been extensively studied[2,3], the manipulation of these excitons required to alter the valley-state or tune the emission energy and intensity is still lacking. Here, we demonstrate such control over interlayer excitons in MoSe$_2$/WSe$_2$ heterostructures. The encapsulation of our stack with h-BN ensures ultraclean interfaces, allowing us to resolve four separate narrow interlayer emission peaks. We observe two main interlayer transitions with opposite helicities under circularly polarized excitation, either conserving or inverting the polarization of incoming light. We further demonstrate control over the wavelength, intensity, and polarization of exciton emission by electrical and magnetic fields. Such ability to manipulate the interlayer excitons and their polarization could pave the way for novel excitonic and valleytronic device applications.**

Electronic devices rely on the manipulation of the charge degree of freedom to store and process information. To overcome their fundamental limitations related to power dissipation, different degrees of freedom could be utilized in new device concepts. Using the electron spin, for example, has been considered as an attractive alternative[4] to charge-based devices. Another possibility consists of manipulating the valley degree of freedom[5,6], corresponding to the minima (maxima) of the conduction (valence) band a carrier is occupying. To be able to store and transmit information using the valley degree of freedom, a material needs at least two inequivalent valleys. In this regard, 2D semiconductors such as transition metal dichalcogenides (TMDCs) are particularly attractive for valley manipulation, as their band structure has K and K´ valleys, which behave as a ½-pseudospin system[7,8]. Neutral and charged excitons in these materials are of high interest since they also possess a valley degree of freedom originating from their constituent electrons and holes and can be optically addressed with circularly polarized light[9–12]. The main challenge is to gain a deeper understanding of the phenomena and to be able to manipulate valley carriers in the desired way. In analogy with present optoelectronic devices, control over the emission intensity and wavelength are essential, as well as the possibility to manipulate the logic state of the system, i.e. the helicity of light. To this purpose, we study interlayer excitons hosted in a dual-gated $WSe_2/MoSe_2$ heterostructure, demonstrating control of these key parameters by application of electric and magnetic fields.

Our device consists of a contacted $MoSe_2/WSe_2$ heterobilayer encapsulated in h-BN, with a graphene bottom gate and a top transparent Pt gate (Figure 1a). The stack is realized by dry-transfer technique[13] on a doped silicon substrate covered with 270 nm of $SiO_2$. This device architecture allows us to perform optical measurements on the sample while applying different voltages through the top and bottom gates, as well as the global back-gate. Figure 1b shows the optical microscopy image of the completed heterostructure device on which we obtain the results presented in the main text.

Stacking MoSe$_2$ and WSe$_2$ on top of each other results in the creation of a van der Waals (vdW) heterostructure,[14] with a type-II band alignment,[15] as schematically depicted in Figure 1c. The valence band minima (VBM) and the conduction band maxima (CBM) reside in WSe$_2$ and MoSe$_2$ respectively, with an ultrafast charge transfer confining electrons and holes to separate layers.[2] We perform µ-PL measurements on our heterostructure, probing each monolayer and heterobilayer in different positions at a temperature of 4 K. A 647 nm continuous wave (CW) laser is used for excitation, with 200 µW incident power (see materials and methods).

The high quality of the h-BN-encapsulated heterostructure allows us to observe bright and sharp photoluminescence peaks from individual monolayers (Figure 1d, insets), with full-width half maxima (FWHM) between 4 and 15 meV (see Supplementary Figure S1). In the heterobilayer region, we observe extreme quenching of the intralayer excitonic peaks, together with the appearance of low-energy emission (Figure 1d and Supplementary Figure S2) around 1.4 eV due to interlayer exciton formation. In contrast with the previous reports[2,16], we clearly resolve two distinct emission peaks (Figure 1e), with FWHM around 5 meV and energy separation of ~25 meV between them, which is a result of the improved heterostructure quality. We will refer to these two features as IX$_1$ and IX$_2$, centered at 1.398 eV and 1.423 eV respectively. Considering the energy difference between IX$_1$ and IX$_2$, the doublet feature could be a result of the spin splitting of the conduction band of MoSe$_2$.[2]

When inspecting in detail, we observe that each peak has an additional small satellite peak around 5 meV lower in energy. We attribute this to defect states near the valence band edge of WSe$_2$, and draw the schematic picture of optical transitions in our system as in Figure 1f. It should be noted that while the straightforward picture of a K-K transition direct in momentum space could be tempting, the exact nature of the transitions from which the interlayer emission originates is still controversial[17]. Proposed models for the two peaks are based on the existence of bright and dark excitons at the K point[2] or an additional momentum-indirect transition[18]. A very recent work interpreted the effect as the result of a transition between VBM in $WSe_2$ at K/K´ points and CBM in $MoSe_2$ at Q/Q´ points with relaxed selection rules[17]. This is also likely the origin of our spectra since $IX_1$ and $IX_2$ have comparable intensities.

Since electrons and holes are confined to separate layers, interlayer excitons have a defined dipole momentum $\vec{p}$ oriented perpendicularly to the heterostructure plane. This creates the opportunity to tune the energy of the interlayer emission by applying an external electric field $\vec{E}$ along the axis of the dipole. In a semi-classical picture, we can then expect an energy shift $\Delta U \sim -\vec{p}\cdot\vec{E}$. To this end, we apply a vertical field at constant carrier concentration, resulting in linear tuning of the interlayer emission energy, symmetric for positive and negative fields. As shown in Figure 2a, a modulation of the IX emission maximum of $\Delta U \sim 138$ meV is obtained, from 1.34 to 1.47 eV. It is interesting to compare these results with a recent work on TMDC bilayers[19], where the observed Stark shift is only related to the magnitude of $\vec{E}$, and not to its direction, since the dipole can reverse its orientation. Here, the orientation is fixed, and thus the direction of the field can induce both positive and negative energy shifts of larger magnitudes. A linear fit of the emission peak as a function of gate voltage yields a tuning rate of ~500 meV·nm·V$^{-1}$. From this value, we can also obtain a qualitative estimation of the dipole size $d \sim \Delta U/qE \sim 0.5$ nm, which is compatible with the expected interlayer spacing between monolayers[20]. It should be noted that the shape and intensity of the spectra do not show significant change as the electric field is modulated at constant doping (see Supplementary Figure S3).

Our device structure also allows control over the emission intensity. For this, we ground the monolayers while applying voltage to the top gate. As a result of this electrostatic doping, a strong modulation in the intensity of the two interlayer peaks is observed (Figure 2b). For negative values of $V_{TG}$, the intensity of the $IX_2$ peak is first reduced, and then suppressed around -4 V. At the same time, $IX_1$ becomes broader and starts to dominate the spectrum. On the contrary, at high positive voltages, we observe that $IX_2$ becomes the dominant emission feature, while $IX_1$ decreases in intensity and becomes quenched at higher electron concentration achieved by dual gating when both top- and bottom- local gates are exploited (see Supplementary Figure S4). This effect can be explained by the filling of the lower conduction band of $MoSe_2$ with increased doping, driving more photoexcited electrons into the upper level. This interpretation is also supported by the observation of a faster increase in the intensity of $IX_2$ with increasing laser power in the absence of electrostatic doping (see Supplementary Figure S5). We assign the kink in the peak position at $V_{TG}$ = 0.75 V to the charge neutrality point. This is also supported by the observation of intralayer trion emission disappearing at this gating configuration (see Supplementary Figure S6).

Excitonic valleytronic devices should have an optical input and output, with the information encoded in the polarization of light. Here, selectively addressing the valley degree of freedom of excitons with polarized light is critically important. To this end, we investigate the polarization-resolved photoluminescence from our heterostructure. Figure 3 shows spectra recorded for left- and right- polarized PL in the case of circularly- and linearly-polarized excitation. As expected, the emission intensity for positive ($\sigma^+$) and negative ($\sigma^-$) helicity are the same in the case of linear excitation. The situation changes with circularly polarized excitation. We observe robust conservation of the incident polarization from monolayer $WSe_2$, but not from $MoSe_2$[21,22] (see Supplementary Figure S7). As shown earlier, our clean interfaces in h-BN-encapsulated heterostructures allow us to resolve the two different optical transitions, $IX_1$ and $IX_2$. Here, we observe that $IX_1$ and $IX_2$ have *opposite* behavior under circularly polarized excitation. As shown in Figure 3a and 3b, $IX_1$ emits prevalently with the same helicity as the pump light, while $IX_2$ has a stronger emission for the opposite circular polarization. It has been reported before that $MoSe_2/WSe_2$ heterobilayers excited with polarized light show PL with the same helicity as the incident radiation[16]. This is due to the broad emission, which hinders resolving the two peaks. In the insets of Figure 3a-c we also report the calculated polarization degree $\rho = \frac{I(\sigma^+) - I(\sigma^-)}{I(\sigma^+) + I(\sigma^-)}$, which shows a similar magnitude and opposite sign for the two peaks. Note that a similar behaviour has not been observed in any other 2D system, and could in principle enable more complex device operation. The presence of spin-flipping transitions can tentatively be explained by considering the relaxed selection rules in the case of indirect interlayer excitons. Indeed, DFT calculations of the matrix elements for recombination between spin-orbit- split conduction band at the Q point and valence band at the K point yield similar weights and opposite helicities[17]. We measure polarization up to 27% for $IX_1$ and -25% for $IX_2$.

While polarization conservation is an interesting property of this system, realizing valleytronics requires implementing logic operations, with logical negation being one of them. We can achieve this in our device using gate modulation, previously demonstrated in Figure 2b. As we have seen, strong electron doping enhances $IX_2$, while at small gate voltages $IX_1$ dominates. Thanks to the opposite polarization of the two peaks, we can then change the device operation between a polarization-inverting and polarization-preserving regime. We show the corresponding results in Figure 4 (see Supplementary Image S8 for raw left and right polarization): both excitonic peaks are clearly visible in the upper (positive) half of the map, with opposite helicity. In Figure 4c, the spectra corresponding to $V_{TG} = 0$ V is presented. Due to the higher intensity of $IX_1$ peak, the total polarization of the integrated signal is positive (i.e., of the same sign as the excitation). This is even more clearly visible in Figure 4f, where the spatial image of the exciton polarization acquired on the CCD is shown. For $V_{TG}$ = +8 V, the situation is reversed, as seen on Figure 4b and 4e. In this configuration, IX$_2$ emission is stronger, resulting in an overall negative value of *ρ* and our device operates here as a polarization inverter. Even more surprising is the behaviour in the p-doped region. As seen before, the higher energy IX$_2$ peak is suppressed at negative gate voltages, so one would expect the device to strongly preserve the helicity when p-doped. On the contrary, IX$_1$ polarization behaviour is now completely reversed (see lower half of Figure 4a). In Figure 4d we show the spectra recorded for $V_{TG}$ = -8 V, demonstrating that the polarization-inverting emission is indeed coming from the lower-energy IX$_1$. Just as in the case of $V_{TG}$ = +8 V, we obtain a globally negative polarization, also visible in Figure 4g.

One possible explanation for the polarisation inversion in our device could be related to the moiré pattern produced by lattice mismatch in vdW heterostructure, which can significantly alter the electronic structure[23]. In particular, recent theoretical work has shown that interlayer excitons have relaxed selection rules, which brightens transitions that would normally be dark in monolayers[7]. Yu et al.[24] have shown that both spin singlet and triplet states (arising due to conduction band splitting) can produce bright transitions coupled with in-plane polarization, with opposite selection rules. These optical selection rules could be reversed with the application of electric fields.[25]

Since the valley pseudospin is associated with a magnetic moment, it is possible to manipulate it with an external magnetic field.[26] Such control could also help us better understand the underlying physics of the phenomenon. Towards this, we perform polarization-resolved photoluminescence measurements in a magnetic field *B*, from -3 T to +3 T. The results are shown in Figure 5. Similarly to the monolayer case[27], the σ$^+$ and σ$^-$ components of the PL peaks experience opposite energy shifts, with a splitting $\Delta E^{IX} = E^{IX}_{\sigma^-} - E^{IX}_{\sigma^+}$ proportional to *B*. Interestingly, the same polarization component for each peak experiences an opposite energy shift, further supporting the appearance of the interlayer doublet (IX$_1$ and IX$_2$) as a result of the conduction band spin splitting. Plotting separately the energy splitting for IX$_1$ and IX$_2$ as in Figure 5a, it is clear that the two transitions show opposite Zeeman effects. For a more quantitative analysis of the data, we extract the peak positions form a Gaussian fit. In Figure 5b we show the calculated shift for IX$_1$ and IX$_2$, $\Delta E^1 = E^1_{\sigma^+} - E^1_{\sigma^-}$ and $\Delta E^2 = E^2_{\sigma^+} - E^2_{\sigma^-}$. A linear fit yields slopes similar in magnitude but opposite in sign, around $\mp 0.5$ meV/T. Using the Zeeman energy shift $\Delta E = g_{\text{eff}} \mu_B B$ where $\mu_B$ is the Bohr magneton and $B$ the applied field, we calculate an effective g-factor $|g_{\text{eff}}| = 8.8 \pm 0.2$. Our result is considerably larger than previously reported effective g-factors in monolayer TMDCs (normally around -4).[28]

In conclusion, we have demonstrated electrical control over the wavelength, intensity and polarization of emission from interlayer excitons hosted in a WSe$_2$/MoSe$_2$ vdW heterostructure. The ability to fine-tune the emitted radiation is key to practical optoelectronics and could pave the way for novel applications for excitonic devices. Our encapsulated device allows probing two interlayer transitions with opposite helicities. Even more importantly, polarisation conservation or reversal is gate-tuneable, enabling for the first time a polarization-inverting action. We measure high and opposite g-factors for both IX$_1$ and IX$_2$ transitions. These results are relevant in the context of valleytronic devices since they enable easy manipulation of the information encoded in the polarization of light.

**ACKNOWLEDGEMENTS**

The authors acknowledge the help of Zdenek Benes of CMi for electron beam lithography. D.U., A.C., A.A. and A.K. would like to acknowledge support by the Swiss National Science Foundation (Grant 153298), H2020 European Research Council (ERC, Grant 682332), and Marie Curie-Sklodowska-Curie Actions (COFUND grant 665667). A.K. acknowledges funding from the European Union's Horizon H2020 Future and Emerging Technologies under grant agreement No 696656 (Graphene Flagship). K.W. and T.T.

acknowledge support from the Elemental Strategy Initiative conducted by the MEXT, Japan and JSPS KAKENHI Grant Numbers JP15K21722 and JP25106006.

## MATERIALS AND METHODS

**HETEROSTRUCTURE FABRICATION**: Few-layer graphene flakes for the bottom gate were obtained by exfoliation from graphite (NGS) on Si/SiO$_2$ substrates and patterned in the desired shape by e-beam lithography and oxygen plasma etching. The heterostructure was then fabricated using polymer-assisted transfer of mono- and few-layer flakes of hBN, WSe$_2$ and MoSe$_2$ (HQ Graphene). Flakes were first exfoliated on a polymer double layer. Once monolayers were optically identified, the bottom layer was dissolved with a solvent and free-floating films with flakes were obtained. These were transferred using a home-built setup with micromanipulators to carefully align flakes on top of each other. Polymer residue was removed with a hot acetone bath. Once completed, the stack was thermally annealed under high vacuum conditions at $10^{-6}$ mbar for 6 h. Finally, electrical contacts were fabricated using e-beam lithography and metallization (80 nm Pd for contacts, 8 nm Pt for the top-gate).

**OPTICAL MEASUREMENTS:** All measurements presented in the work were performed in vacuum at a temperature of 4.2 K. Excitons were optically pumped by a continuous wave (cw) 647 nm laser diode focused to the diffraction limit with a beam size of about 1 µm. The incident power was ~200 µW. Spectral and spatial characteristics of the device emission were analysed simultaneously. The emitted light was acquired using a spectrometer (Andor Shamrock with Andor Newton CCD camera), and the laser line was removed with a long pass 650 nm edge filter. For spatial imaging, we used a long-pass 850 nm edge filter so that the laser light and most of the emission from monolayers were blocked. Filtered light was acquired by a CCD camera (Andor Ixon). For polarization-resolved measurements, a lambda-quarter plate on a rotator together with a linear polarizer were used to select the polarization of incident light. A similar setup was used to image the two polarizations on the CCD camera. To acquire separate spectra for left- and right-circularly polarized light, a lambda-quarter and a displacer were employed.

**AUTHOR CONTRIBUTIONS**

A.K. initiated and supervised the project. A.C. fabricated the device. K.W. and T.T. grew the hBN crystals. D.U. performed the optical measurements, assisted by A.C.. A.C., and D.U. analysed the data with input from A.A. and A.K.. All the authors contributed to the writing of the manuscript.

# FIGURES

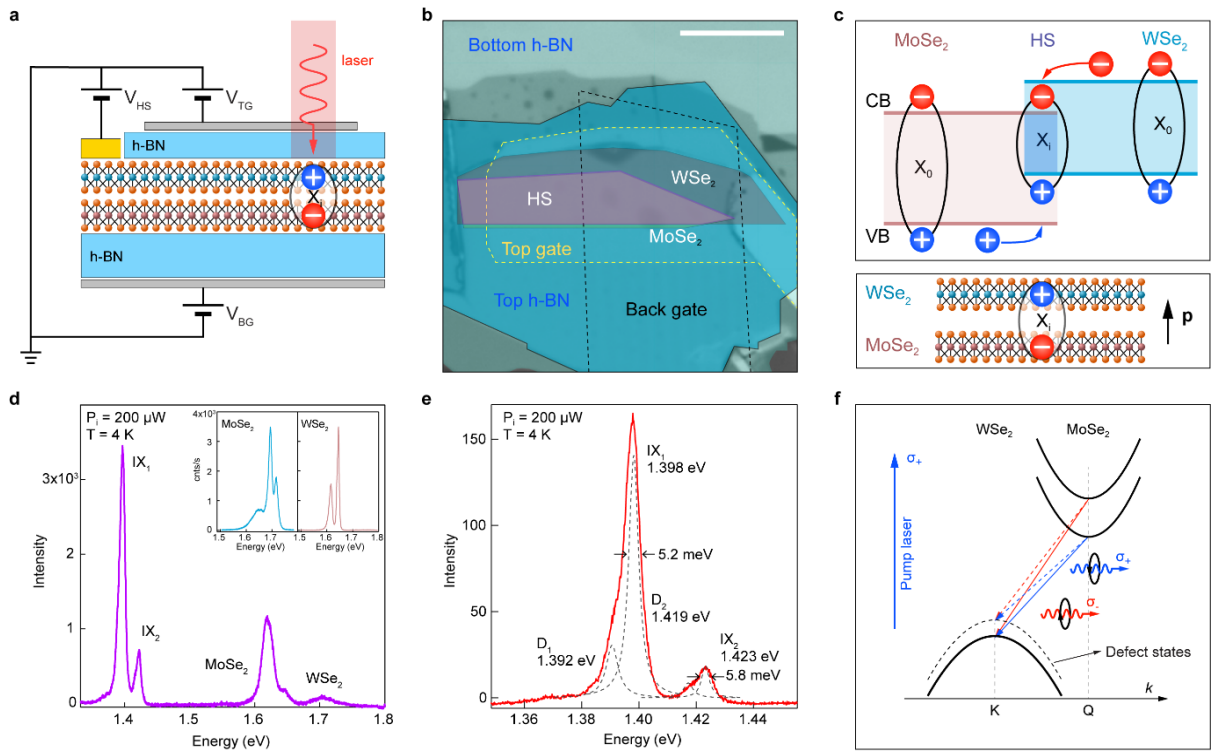

**Fig. 1| Device characterization. a,** schematic depiction of the device structure. **b,** optical image of the device. The pink area indicates the heterobilayer area. Scale bar is 10 µm. **c,** band alignment in MoSe$_2$/WSe$_2$ heterobilayer. **d,** PL spectrum from the heterostructure, showing emission from MoSe$_2$, WSe$_2$ and interlayer excitons. Insets show PL spectra from MoSe$_2$ and WSe$_2$ monolayers. **e,** detail of the PL spectrum from the heterobilayer, with numerical fits for the emission peaks. **f,** schematic of the allowed transitions in the K valley for the structure under consideration.

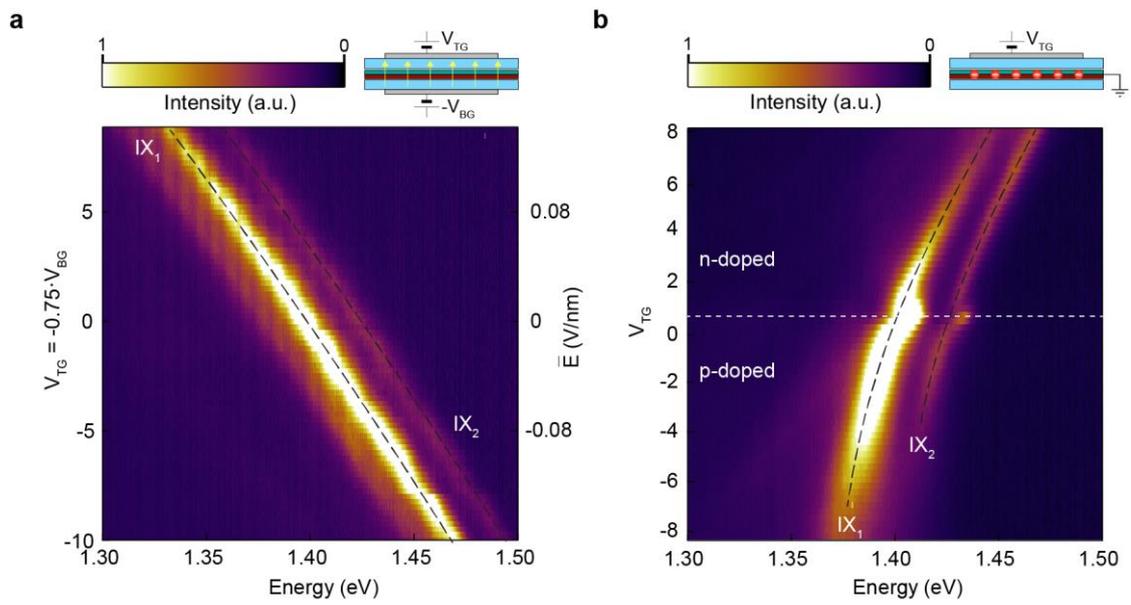

**Fig. 2| Electrical control of interlayer excitons. a,** map of PL emission as a function of applied gate voltages $V_{TG}$ and $V_{BG}$ when sweeping at constant doping. **b,** map of PL emission as a function of gate voltage when electrostatically doping the device.

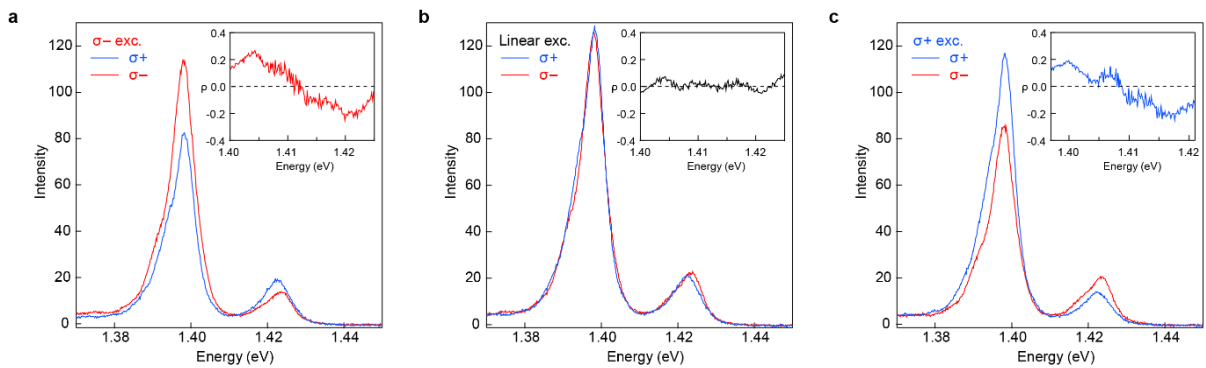

**Fig. 3| Polarization-resolved PL.** µ-PL spectra for left and right emission from the heterobilayer in the case of **a,** left-circularly-polarized excitation. **b,** linear excitation and **c,** right-circularly-polarized excitation

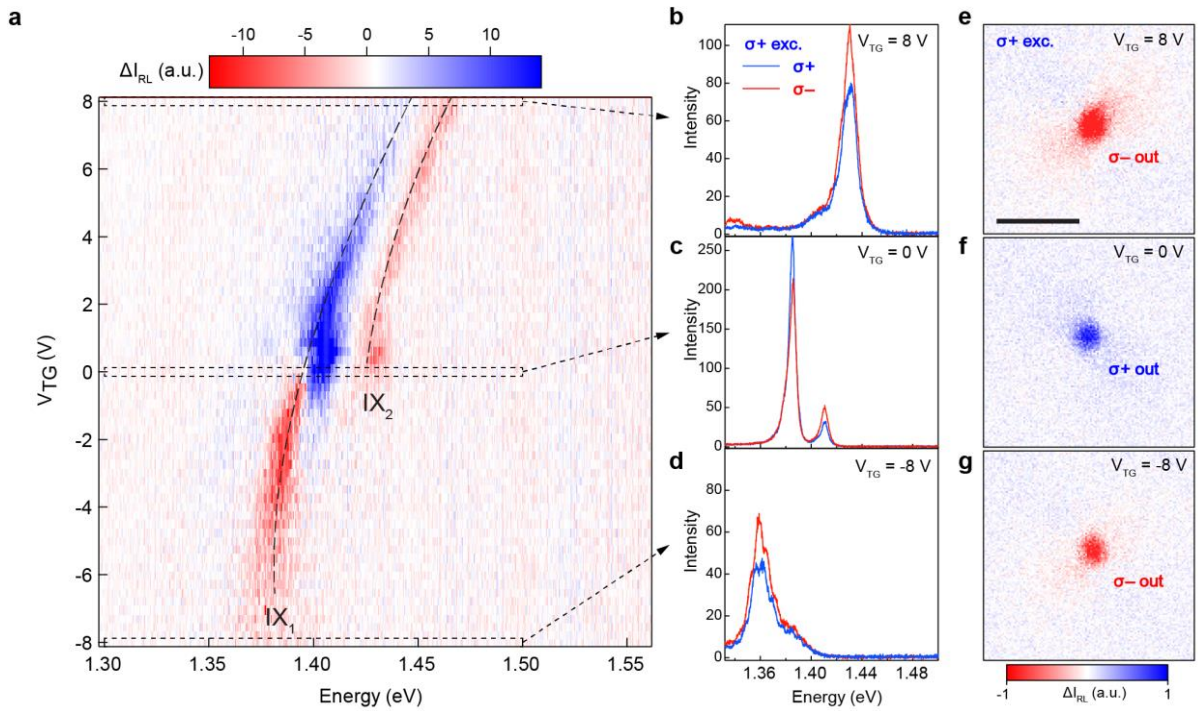

**Fig. 4| Electrical control of polarization. a,** μ-PL map of the difference in right and left emission intensities when device is pumped with right-circularly-polarized light: $\Delta I_{RL} = I_R - I_L$ as a function of the gate voltage $V_{TG}$. **b-d,** PL spectra for $V_{TG}$ = -8 V, 0 V and +8 V. **e-g,** spatial imaging of $\Delta I_{RL}$ at the same gate voltages. Scale bar is 5 μm.

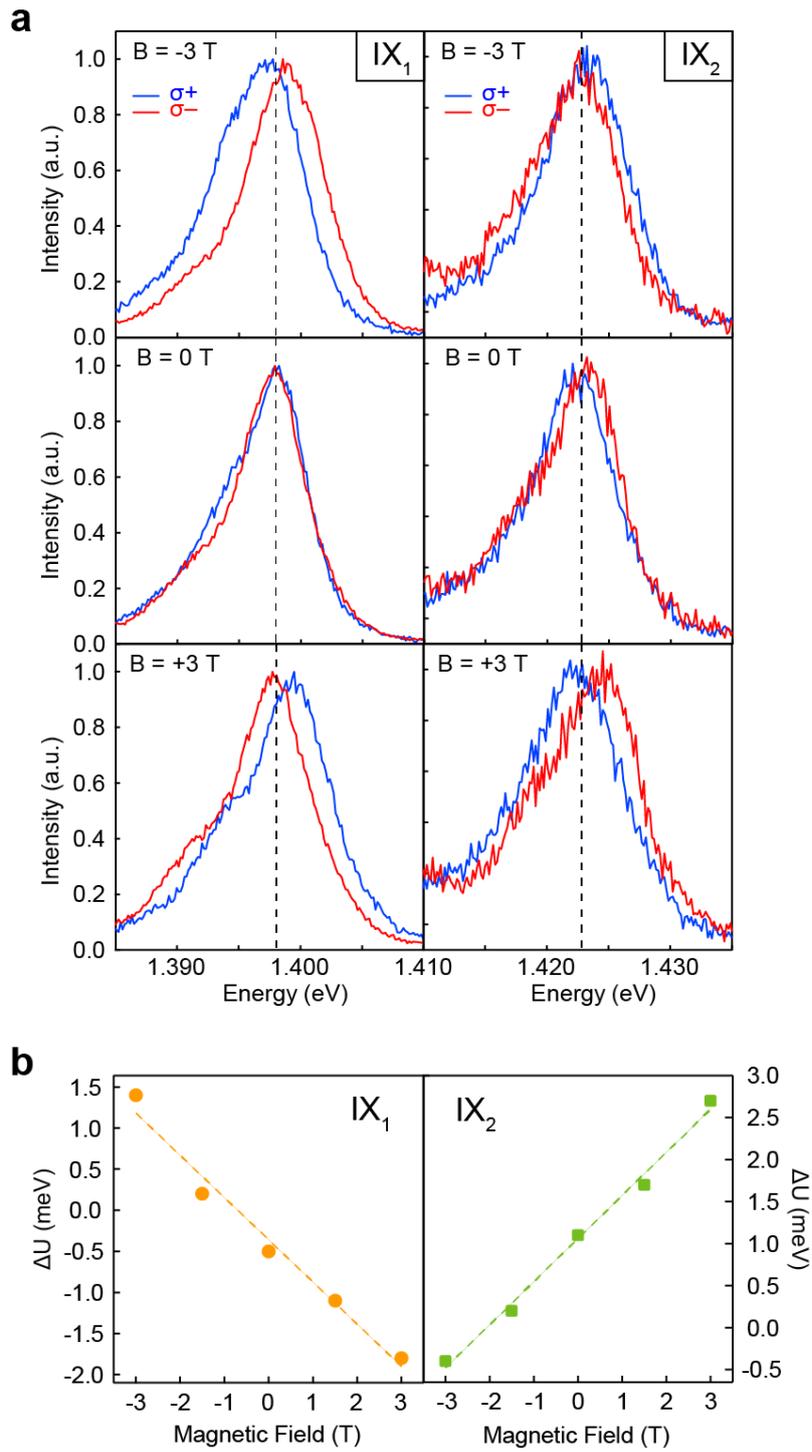

**Fig. 5| Magnetic field effect on interlayer excitons. a,** PL spectra for applied magnetic fields between -3 T and +3 T for left- and right- circularly polarized emission for IX$_1$ (left column) and IX$_2$ (right column). **b,** energy shift ΔU between left- and right-CP peaks for IX$_1$ and IX$_2$.